# Early estimates on monkeypox incubation period, generation time and reproduction number in Italy, May-June 2022

Giorgio Guzzetta (1), Alessia Mammone (2), Federica Ferraro (2), Anna Caraglia (2), Alessia Rapiti (2), Valentina Marziano (1), Piero Poletti (1), Danilo Cereda (3), Francesco Vairo (4), Giovanna Mattei (5), Francesco Maraglino (2), Giovanni Rezza (2)*, Stefano Merler (1)*,#

**Affiliations:**

(1) Center for Health Emergencies, Fondazione Bruno Kessler, Trento, Italy

(2) Directorate General of Health Prevention, Ministry of Health, Italy.

(3) Directorate General for Health, Lombardy Region, Milano, Italy

(4) National Institute of Infectious Diseases Lazzaro Spallanzani, Italy

(5) General Directorate for personal care, health and welfare, Emilia Romagna Region, Bologna, Italy

*joint senior authors

#corresponding author: merler@fbk.eu

## Abstract

We analyzed the first 255 PCR-confirmed cases of monkeypox occurred in Italy in 2022. Preliminary estimates indicate: mean incubation period of 9.1 days (95%CI of the mean: 6.5-10.9); mean generation time of 12.5 days (95%CI of the mean: 7.5-17.3); reproduction number in the MSM community of 2.43 (95%CI: 1.82-3.26).

After the first reports of autochthonous cases of monkeypox (MPX) in Europe at the beginning of May, the World Health Organization and the European Centre for Disease Control have alerted member states to report suspect and/or confirmed cases. Apart from early cases reported in the UK, one of whom had travelled to Nigeria (1), most of the cases were identified in adult males, and especially in men who have sex with men (MSM) (1-4).

**The study**

In Italy, suspected MPX cases fitting the criteria of the WHO case-definition (5) are reported to the surveillance system of the Ministry of Health. Only those cases resulting positive to MPX specific polymerase chain reaction (PCR) were considered as confirmed. Information on the main characteristics of the patients (age, gender, date of symptoms onset, rash and other signs, exposure modality, and travel abroad) were collected.

Up to July 8$^{th}$, 2022, a total number of 255 PCR confirmed cases were reported in Italy (Figure 1). All except two were males, and 190 men out of 200 (95%) for which the information was disclosed reported having sex with men; the median age was 37 years (range: 20 - 71 years).

For 139 out of 184 cases for whom the information was available, the rash was localized at the genital/perianal area. Fever was reported in 151 out of 222 cases for whom this information was available.

Information about travelling was available for 228 cases; 86 (37.7%) had travelled abroad, and 25 of these 86 (29.1%) had spent a vacation period on the Canary Islands, suggesting the occurrence of a major amplifying event (Table 1). Only one case had travelled to West Africa and was symptomatic at arrival in Italy.

We estimated the incubation period from 30 cases with known date of symptom onset and for which epidemiological investigations allowed the identification of their likely period of exposure (exact date for 15 cases; dates of sojourn in Canary Islands for 15 cases). The generation time (i.e. the time elapsed between the date of exposure of a confirmed case and those of his secondary cases) was estimated by considering 16 infector-infectee pairs, as identified during contact tracing operations. The two periods were assumed to be distributed as a gamma function and they were estimated following a Bayesian approach similar to what was adopted in (6). Likely dates of exposure were considered for each confirmed case within a Markov chain Monte Carlo (MCMC) procedure. For the estimation of the generation time, we assumed no pre-symptomatic transmission. Therefore, sampling of candidate dates of exposure was repeated if the date of exposure for the infectee was earlier than the date of symptom onset for the infector. Estimates of the generation time were used to compute the net reproduction number. More details are provided in the Technical Appendix.

The mean incubation period was estimated to be 9.1 days (95%CI of the mean: 6.5-10.9; 5$^{th}$ and 95$^{th}$ percentiles of the distribution: 1-24). The mean generation time was estimated to be 12.5 days (95%CI of the mean: 7.5-17.3; 5$^{th}$ and 95$^{th}$ percentiles of the distribution: 4-26). By assuming a mean generation time of 12.5 days and importation from Canary Islands, the mean net reproduction number (mean number of cases generated by a single index case) during the first week of June was estimated at 2.43 (95%CI 1.82-3.26). After June 12, 2022 a progressive decrease of the reproduction number was estimated.

**Discussion**

The first large outbreak of MPX outside Africa is to some extent unique. The analysis of virus genome strongly suggests that the epidemic is caused by the West African clade of the MPX virus (7); however, with the exception of two men who reported travels to Western African countries (1, 8), at least 60% of cases were locally acquired. A retrospective investigation then indicated that, both in Portugal and in the UK, symptoms onset of the first cases dated back at least to April 2022. The presence of skin lesions at the point of sexual contact is suggestive of sexual transmission (9).

Following the early reports of this multi-country outbreak, the Italian Ministry of Health issued a series of recommendations, including case-notification, protective measures reducing contacts and possible exposure to droplets for health care workers, contact tracing with self-surveillance of close contacts, and the possibility of implementing quarantine measures to be decided by local health authorities in particular epidemiological and/or environmental contexts (10).

In Italy, after the first four cases in MSM who had travelled abroad (4), other cases were notified to the Ministry of Health from several Regions (8). The number of cases by date of symptom onset increased over time, particularly in two Italian regions (Lombardy and Lazio, where Milan and Rome are located), with notifications from 11 regional health authorities in total (out of 21). Almost all cases continued to be reported among men. A significant fraction of cases (38.9%) identified in Italy so far has a history of travel abroad, and direct or sexual (mostly male-to-male) contact is still likely to be the main transmission mode. Whether the infection was transmitted through direct contact with skin lesions or body fluids remains undefined. The link with different geographic areas (i.e., Europe and West Africa) underlines the possibility of multiple independent introductions of the virus, suggesting a wide spread of the infection in Western Africa, dating back to the pre-pandemic years (3, 11).

Using a limited number of cases, we provided estimates of the mean incubation period (about 9 days, n=15 individuals with known date of exposure and 15 individuals with known dates of sojourn in Canary Islands) and of the mean generation time (about 12 days, n=16 infector-infectee pairs). Based on the estimated mean generation time, we found that the reproduction number for this outbreak is approximately 2.4, although with a broad uncertainty (1.82-3.26) due to the still limited number of locally acquired confirmed cases. Small variations in the estimated reproduction number (mean values ranging from 2.08 to 2.70) were found when considering different distributions of the generation time (mean 7.5 and mean 17.3 days) and when exploring alternative assumptions on the importation of cases (see Technical Appendix). Our estimates of the reproduction number should be considered as referring to the community of MSM in which MPX is currently spreading and not as an estimate at the general population level. It is unclear to what extent the decrease of the reproduction number estimated after June 12, 2022 results from a reduction of the transmission (e.g., led by an increasing awareness about the risks of infection) and/or from the analysis of incomplete data due to diagnostic and reporting delays. However, considering that the overwhelming majority of cases seems to have been transmitted via sexual contacts, the reproduction number is likely below threshold in the general population. Beside the limited number of cases, our estimates may be biased by several factors: the assumption that cases returning from Canary Islands acquired the infection there, possible recall bias for the dates of exposure, selection bias in the reconstructed infector-infectee pairs (e.g., a recent sexual partner may be more likely to be identified).

Maintaining a high level of public attention and providing non-stigmatizing information to at risk population groups are key factors to contain the spread of MPX virus, also considering the seasonal intensity of aggregation events and recreational activities.

**Acknowledgements**

The authors would like to thank for their collaboration: Corrado Cenci, Arianna Bruscolotti, Andreina Pagini and Ivana Raccio of Directorate General of Health Prevention, Ministry of Health; Francesca Zanella, Francesca Russo, Chiara Pasqualini, Marcello Tirani, Federica Attanasi, Marino Faccini, Gian Luigi Belloli, Giulio Matteo, Gabriella De Carli, Emanuela Balocchini, Daniela

Senatore, Cristina Zappetti, Maria Grazia Zuccali, Domenico Martinelli, of Regional Health authorities; and all healthcare workers in the Regions for notifying the cases.

**References**

1) Vivancos R, Anderson C, Blomquist P, Balasegaram S, Bell A, Bishp L, et al. Community transmission of monkeypox in the United Kingdom, April to May 2022. Euro Surveill 2022;27(22):pii=2200422.
2) Perez Duque M, Ribeiro S, Vieira Martins J, Casaca P, Pinto Leite P, Tavares M, et al. Ongoing monkeypox virus outbreak, Portugal, 29 April to 23 May 2022. Euro Surveill 2022:27(22):pii=2200424.
3) European Centre for Disease Prevention and Control (ECDC). Monkeypox multicountry outbreak-23 May 2022. ECDC: Stockholm; 2022.
4) Antinori A, Mazzotta V, Vita S, Carletti F, Tacconi D, Lapini LE, et al. Epidemiological, clinical and virological characteristics of four cases of monkeypox support transmission through sexual contact, Italy, May 2022. Euro Surveill 2022; 27(22):pii=2200421.
5) WHO, Disease Outbreak News 21/05/2022. Multi-country monkeypox outbreak in non-endemic countries https://www.who.int/emergencies/disease-outbreak-news/item/2022-DON385
6) Miura F, van Ewijk CE, Backer JA, Xiridou M, Franz E, Op de Coul E, Brandwagt D, van Cleef B, van Rijckevorsel G, Swaan C, van den Hof S, Wallinga J. Estimated incubation period for monkeypox cases confirmed in the Netherlands, May 2022. Euro Surveill. 2022 Jun;27(24):2200448. doi: 10.2807/1560-7917.ES.2022.27.24.2200448. PMID: 35713026; PMCID: PMC9205160.
7) Isidro J, Borges V, Pinto M, Ferreira R, Sobral D, Nunes A, et al. First draft genome sequence of Monkeypox virus associated the suspected multi-country outbreak, May 2022. Available from: https://virological.org/t/first-draft-genome-sequence-of-monkeypox-virus-associated-with-the-suspected-multi-country-outbreak-may-2022-confirmed-case-in-portugal/799.
8) Ferraro F, Caraglia A, Rapiti A, Cereda D, Vairo F, Mattei G, Maraglino F, Rezza G. Multiple introductions of MPX in Italy from different geographic areas. Euro Surveill 2022;27(23):pii=2200456. https://doi.org/10.2807/1560-7917 .

9) Otu A, Ebenso B, Walley J, Barcelo JM, Ochu CL. Global human monkeypox outbreak: atypical presentation demanding urgent public health action. Lancet 2022. https://doi.org/10.1016/S2666-5247(22)00153-7 .

10) Ministero della Salute. https://www.salute.gov.it/portale/news/p3_2_1_1_1.jsp?lingua=italiano&menu=notizie&p=dalministero&id=5907

11) Rezza G. Emergence of monkeypox in west Africa. Lancet Infect Dis 2019; 19: 797-798.

**Tables and figures**

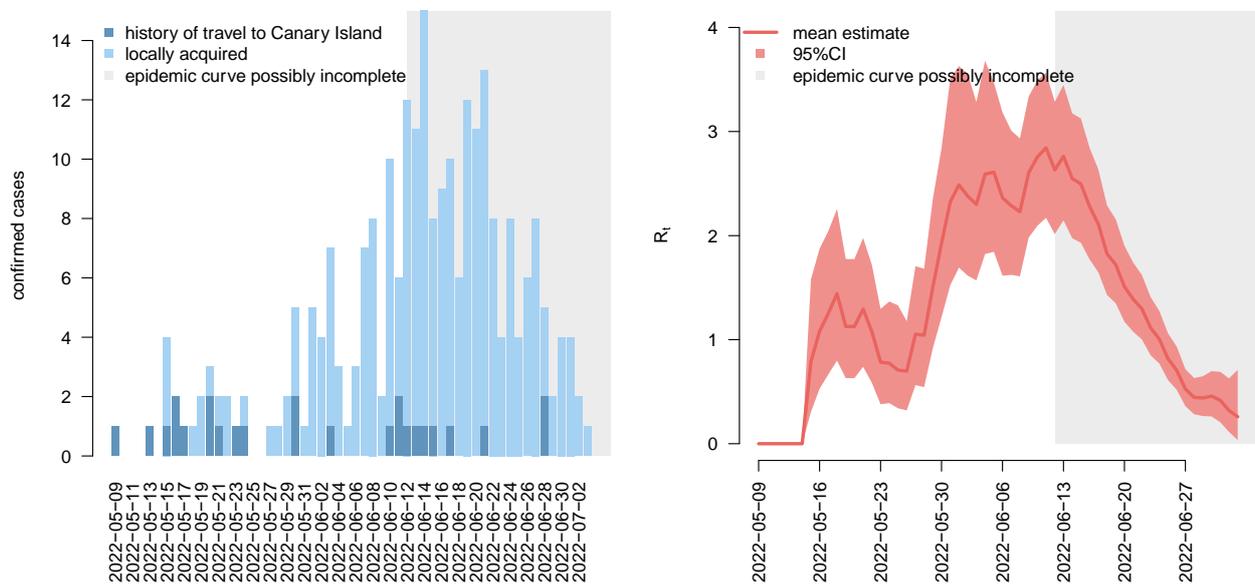

**Figure 1. Epidemic curve and reproduction number**. Left: Number of cases by date of symptom onset and history of travel in Canary Islands. For 4 individuals, the date of symptom onset was unknown. Right: estimate of the net reproduction number over time, estimated from the epidemic curve by date of symptom onset. We assumed that all cases with a history of travel in Canary Islands are imported and that all the others are locally transmitted, and we used a generation time distribution with mean 12.5 days. The gray rectangle identifies the part of the epidemic curve that is possibly incomplete due to diagnostic and reporting delays.

**Table 1.** Characteristics of 255 confirmed MPX cases reported in Italy, up to July 8$^{th}$ 2022.

| Gender | M/F | M % |
|---|---|---|
|  | 253/2 | 99.2 |
| **Age (in years)** | **Median** | **Range** |
|  | 37 | 20-71 |
| **Clinical symptoms** | **n /N*** | **%** |
| Fever | 151/222 | 68.0 |
| Rash | 248/251 | 98.8 |
| Genital/perianal rash | 139/184 | 75.5 |
| **Travels*** | **n /N** | **%** |
| Travel abroad during the last 21 days | 86/228 | 37.7 |
| Travel to Canary Islands during the last 21 days | 25/142 | 17.6 |

* n/N represents the number of positive answers n over the total available answers N

# Technical appendix

# Early estimates on monkeypox incubation period, generation time and reproduction number in Italy, May-June 2022

Giorgio Guzzetta (1), Alessia Mammone (2), Federica Ferraro (2), Anna Caraglia (2), Alessia Rapiti (2), Valentina Marziano (1), Piero Poletti (1), Danilo Cereda (3), Francesco Vairo (4), Giovanna Mattei (5), Francesco Maraglino (2), Giovanni Rezza (2)*, Stefano Merler (1)*,#

**Affiliations:**

(1) Center for Health Emergencies, Fondazione Bruno Kessler, Trento, Italy

(2) Directorate General of Health Prevention, Ministry of Health, Italy.

(3) Directorate General for Health, Lombardy Region, Milano, Italy

(4) National Institute of Infectious Diseases Lazzaro Spallanzani, Italy

(5) General Directorate for personal care, health and welfare, Emilia Romagna Region, Bologna, Italy

*joint senior authors

#corresponding author: merler@fbk.eu

### S1. Incubation period

To compute the incubation period, we considered all cases confirmed up to July 8[th], 2022, with a known date of symptom onset and for which at least one of the following information was reported:

- the latest day of exposure, as ascertained by contact tracing investigation (this information is available for 15 cases);
- a history of travel to Canary Islands, together with the dates of sojourn (this information is available for 15 cases).

As a baseline, we thus considered a total of 30 confirmed cases with prior information on the possible date of exposure.

We assumed that the incubation period is distributed as a gamma function and we estimate the shape ($k$) and scale parameters ($\theta$) following a Bayesian approach similar to the one adopted in (1), based on a Monte Carlo Markov Chain (MCMC) procedure and Metropolis-Hastings sampling. At each MCMC iteration:

- we sampled one date of infection for each confirmed case, considering the information available for each subject. For cases with a known latest day of exposure, the date of infection was fixed to this value; for cases with a travel history to Canary Islands, the sampled date of infection was constrained to be between the date of departure to and return from Canary islands.
- for each confirmed case, we computed the incubation periods as the difference between the date of symptom onset and the sampled date of infections.
- we sampled one value for the shape ($k$) and one value for the scale parameter ($\theta$) of the gamma function.
- we compute the gamma likelihood of observing the computed incubation periods given parameters {$k,\theta$}

The resulting estimates of the parameters are reported in Table S1. The cumulative density function estimated for the incubation period is shown in Figure S1 along with the cumulative distribution of incubation periods associated to the parameter set with the maximum likelihood.

As sensitivities, we estimated the incubation period by considering separately the 15 cases with a known latest date of exposure (sensitivity A) and the 15 cases with a history of travel to Canary Islands (sensitivity B).

**Table S1. Parameters of the incubation period distributions estimated using different methods**

| Method | Baseline (n=30) | Sensitivity A (n=15) | Sensitivity B (n=15) |
| --- | --- | --- | --- |
| Shape (95%CI) | 2.42 (1.26-3.62) | 3.55 (1.57-6.38) | 1.43 (0.62-2.96) |
| Scale (95%CI) | 3.75 (2.17-7.13) | 2.67 (1.21-5.73) | 7.47 (2.97-16.84) |
| Mean (95%CI) (days) | 9.1 (6.5-10.9) | 9.5 (6.3-11.2) | 10.7 (5.5-14.5) |
| 5$^{th}$ and 95$^{th}$ percentiles of the distribution (days) | 1-24 | 2-22 | 1-34 |

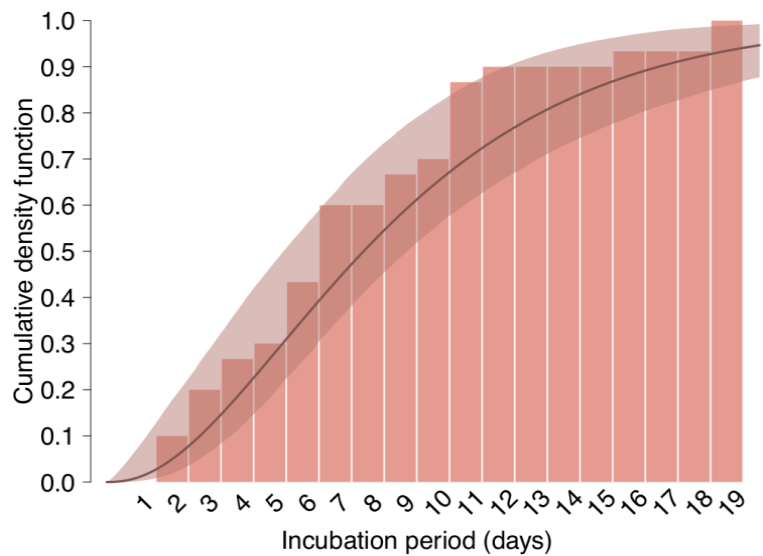

**Figure S1.** Cumulative density function of the incubation period of monkeypox as estimated from 30 cases confirmed in Italy between May and June 2022 (mean, solid line; 95% CI, shaded areas). Bars represent the cumulative distribution of the incubation periods as obtained for these 30 confirmed cases, according to the maximum likelihood parameter set.

**S2. Generation time**

The generation time is defined as the difference between the date of infection of a confirmed case and those of his secondary cases. To estimate the generation time, we followed a MCMC procedure similar to the one used for the incubation period, by assuming a gamma distributed generation time. In the data, there were 16 identified pairs of infector-infectee with known dates of symptom onset. For two cases, an exact date of last exposure was identified during epidemiological investigations and therefore assumed as infection date. For one case reporting traveling to Canary Islands without any other known exposure, we constrained the sampled infection date to be within the dates of departure to and return from Canary Islands. For other cases, the dates of infection were randomly sampled assuming that pre-symptomatic transmission is not possible, i.e., the date of infection for the secondary cases should always be greater or equal to the date of symptom onset of the infector. Dates of infection were sampled using as a uniform prior between values lying within the 95%CI of the incubation period estimated in the baseline analysis.

The obtained distribution of the generation time is reported in Figure S2, and the values are summarized in Table S2.

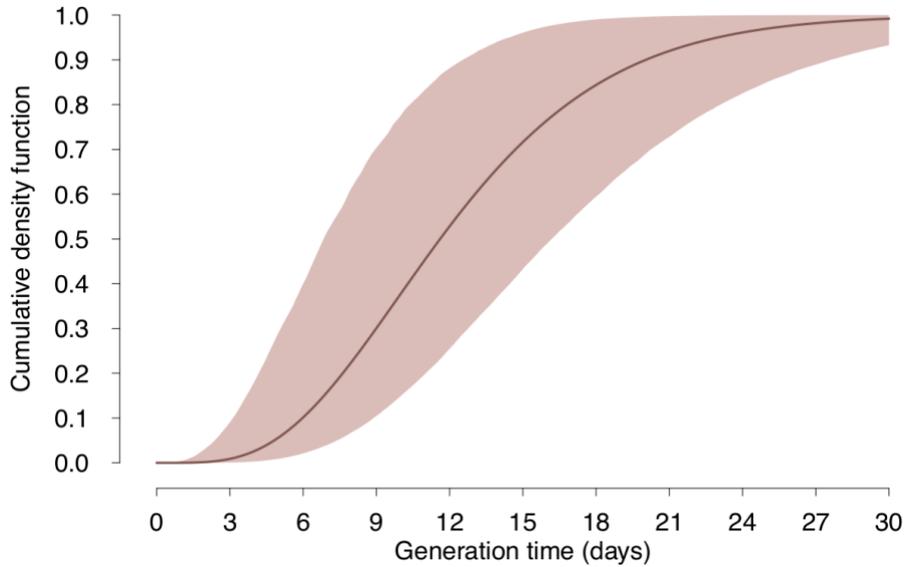

**Figure S2.** Cumulative density function of the generation time of monkeypox as estimated from 16 infector-infectee pairs identified during contact tracing operations conducted in Italy between May and June 2022 (mean, solid line; 95% CI, shaded areas).

**Table S2.** Parameters of the generation time distributions.

|  | Generation time (n=16) |
|---|---|
| **Shape (95%CI)** | 4.85 (3.07 -7.20) |
| **Scale (95%CI)** | 2.57 (1.34 -4.07) |
| **Mean (95%CI) (days)** | 12.5 (7.5-17.3) |
| **5th and 95th percentiles of the distribution (days)** | 4-26 |

## S3. Reproduction number $R_t$

We computed the net reproduction number $R_t$ for the monkeypox outbreak using the epidemic curve of cases by date of symptom onset and the estimated distribution of the generation time, by applying a standard statistical method based on the renewal equation. The posterior distribution of $R_t$ can be computed by applying a Markov Chain Monte Carlo algorithm to the following likelihood function:

$$\mathcal{L} = \prod_{t=1}^{T} P\left( C(t) - I(t); R_t \sum_{s=1}^{T} \varphi(s) C(t-s) \right)$$

Where:

- P(k; λ) is the probability mass function of a Poisson distribution (i.e., the probability of observing k events if these events occur with rate λ).
- C(t) is the total daily number of new cases having symptom onset at time t;
- I(t) is the total daily number of new cases that are not locally transmitted;
- Rt is the net reproduction number at time t to be estimated;
- $\varphi(s)$ is the probability distribution density of the generation time discretized by day, evaluated at days s.

We considered three different assumptions for imported cases:

1) we considered only the case with earliest symptom onset as imported case. This is a rather unrealistic hypothesis that can give an upper bound to the transmissibility of monkeypox;
2) we considered as imported cases all cases with a history of travels to Canary Islands; this assumption was considered as a baseline for the results in the main text;
3) we considered as imported cases all cases with a history of travel abroad; this is also an unrealistic hypothesis that can be considered as a lower bound to the transmissibility of monkeypox.

Cases for which there are no information for travel abroad are always assumed to be locally transmitted.

In Figure S3, we report estimates of the reproduction numbers for the three hypotheses (rows) as obtained by considering the mean and the lower and upper limits of the 95% CI of the generation time, namely 12.5 days, 7.5 days and 17.3 days (columns). Table S3 reports the values of the average reproduction number during the first week of June, i.e., the last week for which the epidemic curve can be reasonably assumed to not suffer from diagnostic and reporting delays.

**Table S3**. Values of the net reproduction number (mean and 95%CI) in the week June 1-7 as obtained by considering the mean (first column); the lower (central column) and the upper (right column) limits of the 95% CI of the generation time. Different rows correspond to different assumptions on importation of cases.

| Generation time ($T_g$) | $T_g$=12.5 days | $T_g$=7.5 days | $T_g$=17.3 days |
|---|---|---|---|
| **Single importation** | 2.51 (1.90 -3.18) | 2.34 (1.77-3.00) | 2.70 (2.04-3.37) |
| **Importations from Canary Islands - baseline** | 2.43 (1.82-3.26) | 2.24 (1.71-2.99) | 2.63 (1.96-3.41) |
| **Importations from abroad** | 2.23 (1.61-2.92) | 2.08 (1.50-2.80) | 2.41 (1.77-3.16) |

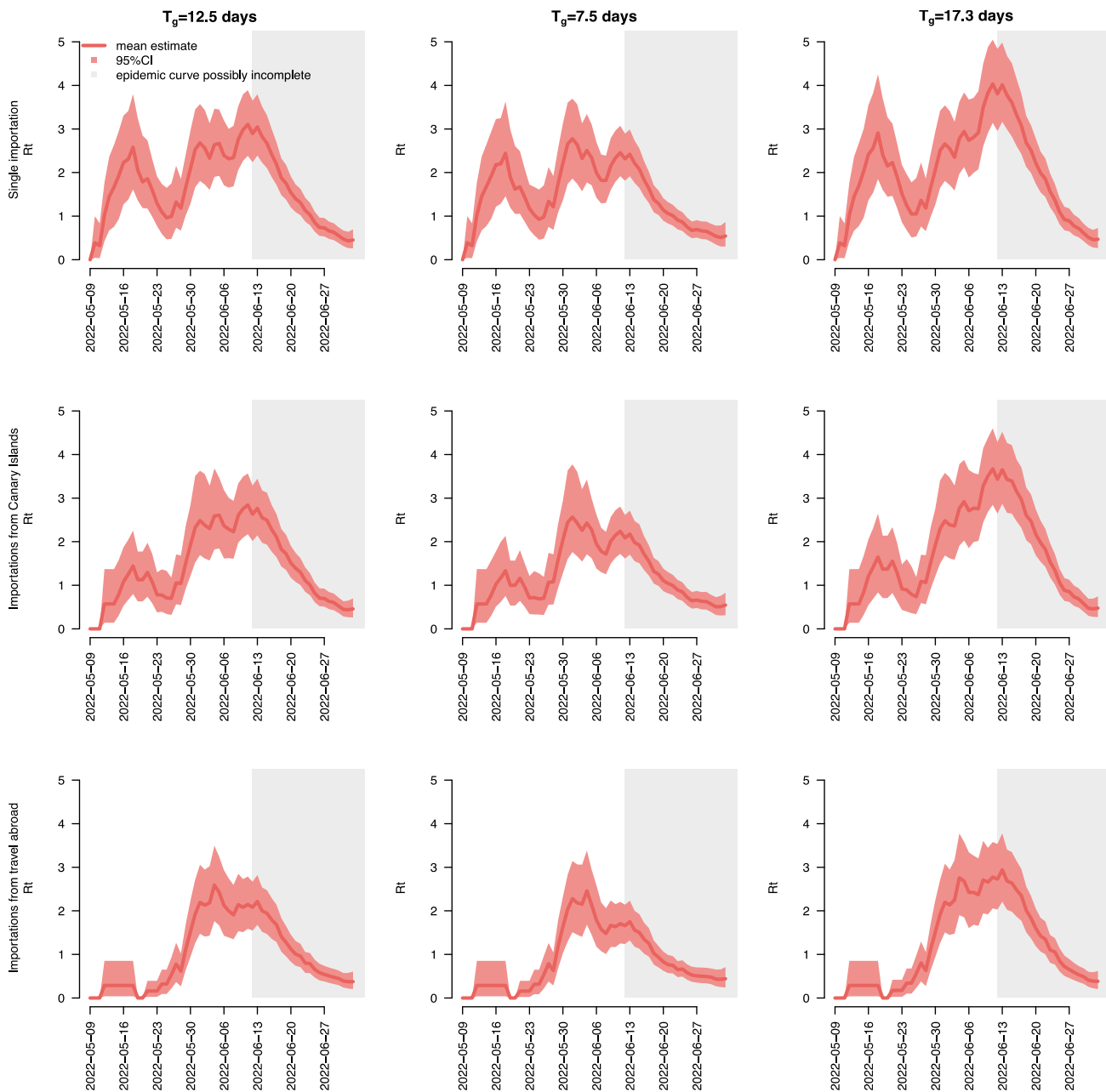

**Figure S3.** Net reproduction number (mean and 95%CI) under the assumption that only the first case was imported (top row), that all cases with a history of travel to Canary Islands were imported (middle row), or that all cases with a history of travel abroad were imported (bottom row). Different distributions of the generation time were used: the first distribution (left column) has mean 12.5 days (corresponding to mean estimates reported in Table S2); the second distribution (centre column) has mean 7.5 days (corresponding to the lower bound of the 95% CI provided in Table S2); the third distribution (right column) has mean 17.3 days (corresponding to the upper bound of the 95% CI provided in Table S2). The baseline results for the main text are those assuming importation from Canary Islands and a mean generation time of 12.5 days (middle row, left column).

**References**

(1) Miura F, van Ewijk CE, Backer JA, Xiridou M, Franz E, Op de Coul E, Brandwagt D, van Cleef B, van Rijckevorsel G, Swaan C, van den Hof S, Wallinga J. Estimated incubation period for monkeypox cases confirmed in the Netherlands, May 2022. Euro Surveill. 2022 Jun;27(24):2200448.